\documentclass[sigconf]{acmart}

\usepackage{amsfonts,amssymb,booktabs}
\usepackage{amsmath}
\usepackage{float}
\usepackage{svg}
\usepackage{graphicx}
\usepackage[ruled,linesnumbered,noend]{algorithm2e}
\usepackage{algpseudocode}
\usepackage{colortbl}
\usepackage{pifont}
\usepackage{utfsym}
\usepackage{fontawesome}
\usepackage{setspace}
\usepackage{subfig}
\usepackage{balance}
\usepackage{makecell}
\usepackage{stfloats}
\usepackage{multirow}

\usepackage{hyperref}
\hypersetup{
	colorlinks=true,
	linkcolor=black,
	filecolor=blue,      
	urlcolor=black,
	citecolor=black,
}

\usepackage{color}

\newcommand{\SystemName}{HiNa-MoE\xspace}
\newcommand{\tmm}[1]{{\small\texttt{tmm#1}}}
\newcommand{\load}[1][]{%
  {\small\texttt{tileload#1}}%
}
\newcommand{\cparagraph}[1]{\vspace{1.5mm}\noindent\textbf{#1}}

\newcommand{\cmark}{\text{\checkmark}} 
\newcommand{\xmark}{\text{$\times$}}     
\newcommand{\kt}{KTransformers\xspace}

\usepackage{etoolbox}
\makeatletter
\patchcmd{\authornote}{\g@addto@macro\addresses{\@authornotemark}}{}{}{}
\makeatother

\makeatletter
\AtBeginDocument{%
  \def\ps@plain{%
    \let\@mkboth\@gobbletwo
    \def\@oddhead{}%
    \def\@oddfoot{}%
    \def\@evenhead{}%
    \def\@evenfoot{}%
  }%
}
\makeatother

\copyrightyear{2026}\acmYear{2026}
\acmConference[PACT '26]{International Conference on Parallel Architectures and Compilation Techniques}{October 19-22, 2026}{Chicago, IL, USA}
\acmBooktitle{International Conference on Parallel Architectures and Compilation Techniques (PACT '26), October 19-22, 2026, Chicago, IL, USA}
\acmDOI{10.1145/3838684.3846846}
\acmISBN{979-8-4007-2915-7/2026/10}

\AtBeginDocument{%
  }

\begin{document}
\begin{sloppypar}

\title{\SystemName: High-Performance, Non-Intrusive MoE Inference on CPUs with Matrix Engines}

\author{Weiling Yang}
\authornotemark[1]
\authornote{Equal Contribution}
\affiliation{%
  \institution{College of Computer Science and Technology, National University of Defense Technology}
  \city{Changsha}
  \country{China}
}
\email{w.yang@nudt.edu.cn}

\author{Junwen Zhang}
\authornotemark[1]
\affiliation{%
  \institution{College of Computer Science and Technology, National University of Defense Technology}
  \city{Changsha}
  \country{China}
}
\email{zhang_junwen@nudt.edu.cn}

\author{Dezun Dong}
\authornote{Corresponding author}
\authornotemark[2]
\affiliation{%
  \institution{College of Computer Science and Technology, National University of Defense Technology}
  \city{Changsha}
  \country{China}
}
\email{dong@nudt.edu.cn}

\author{Jianbin Fang}
\authornotemark[2]
\affiliation{%
  \institution{College of Computer Science and Technology, National University of Defense Technology}
  \city{Changsha}
  \country{China}
}
\email{j.fang@nudt.edu.cn}

\author{Enda Yu}
\affiliation{%
  \institution{College of Computer Science and Technology, National University of Defense Technology}
  \city{Changsha}
  \country{China}
}
\email{yuenda@nudt.edu.cn}

\author{Zhe Bai}
\affiliation{%
  \institution{College of Computer Science and Technology, National University of Defense Technology}
  \city{Changsha}
  \country{China}
}
\email{baizheo3o@nudt.edu.cn}

\author{Xiaopeng Deng}
\affiliation{%
  \institution{College of Computer Science and Technology, National University of Defense Technology}
  \city{Changsha}
  \country{China}
}
\email{dengxiaopeng25@nudt.edu.cn}

\renewcommand{\shortauthors}{Weiling Yang et al.}

\begin{abstract}



Mixture-of-Experts (MoE) inference is increasingly deployed in local and on-premise environments, where expert parameters often exceed GPU memory capacity. In latency-sensitive, low-concurrency settings, repeatedly staging routed-expert weights from CPU memory to the GPU can be prohibitive, leaving routed-expert feed-forward networks (FFNs) on the critical path of multi-socket CPUs.
Existing CPU accelerations often rely on intrusive, hardware- or topology-specific requirements, such as AMX-specific weight layouts or manual NUMA-aware placement. These requirements reduce portability and complicate integration with standard CPU--GPU offloading pipelines.
We present \SystemName, a high-performance, non-intrusive operator library for MoE inference on CPUs with matrix engines. \SystemName (1) exploits AMX with an optimized micro-kernel that keeps expert weights in standard layouts and instead fuses lightweight layout transforms into token gathering and stores; (2) applies NUMA-aware task partitioning that minimizes cross-socket traffic under a simple page-interleaved policy without modifying the framework allocator; and (3) converts decode-phase memory-bound matrix--vector operations into small matrix--matrix execution to better utilize AMX. 
Across multiple server-class CPUs and real-world MoE models, \SystemName achieves up to 3.37$\times$ speedup for FFN kernels and up to 2.09$\times$ end-to-end inference speedup over state-of-the-art baselines, while remaining plug-and-play with existing frameworks and deployment workflows.

\end{abstract}


\keywords{Mixture-of-Experts, CPU Inference, NUMA Optimization, AMX, Non-Intrusive Design}



\maketitle

\section{Introduction}\label{sec:intro}

Mixture-of-Experts (MoE) large language models (LLMs) offer an appealing quality--efficiency trade-off by using conditional computation: each token activates only a small subset of experts in the feed-forward network (FFN) layers~\cite{DBLP:journals/jmlr/FedusZS22,DBLP:journals/corr/abs-2401-04088,DBLP:journals/corr/abs-2412-19437,DBLP:journals/corr/abs-2505-09388}.
This sparsity reduces arithmetic work relative to dense LLMs while preserving high model capacity, and it has become a common design point in production models such as Mixtral~\cite{MistralAI}, Qwen-MoE~\cite{qwen}, and DeepSeek~\cite{ds3}.

\noindent\textbf{Problem scope.}
In local and on-premise deployments, MoE inference increasingly runs on CPU--GPU nodes with limited GPU memory~\cite{vllm_cpu,DBLP:conf/ics/YuDZBYWLWL26,DBLP:conf/asplos/FangHHLC0YZ25,DBLP:conf/iclr/KamahoriTGZK25,DBLP:conf/sosp/ChenXZTWDCYLQZO25,IPEX,DBLP:conf/icpp/ZhangYYFZD26}.
Since the complete set of expert parameters can exceed GPU memory capacity, practical systems often keep some or all routed experts in CPU DRAM and execute different experts on the CPU and GPU~\cite{DBLP:conf/sosp/ChenXZTWDCYLQZO25,DBLP:conf/iclr/KamahoriTGZK25,IPEX,DBLP:conf/ics/YuDZBYWLWL26}.
Thus, MoE inference in this setting is inherently a heterogeneous CPU--GPU execution problem: depending on the inference workload, a routed expert may be executed either on the CPU or on the GPU.

\begin{figure}[!t]
    \centering
    \includegraphics[width=0.58\linewidth]{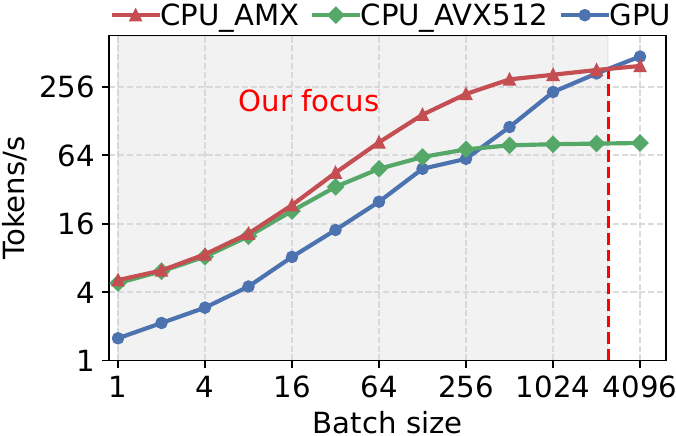}
    \vspace{-1mm}
    \caption{Decoding throughput of CPU- and GPU-based FFN execution for Mixtral-8x7B across batch sizes. CPU execution remains highly competitive over a broad batch regime.
    Results are measured on an Intel Xeon Gold 6430 and an NVIDIA A6000.}
    \label{fig:intro}
\vspace{-3mm}
\end{figure}

A natural strategy is to execute routed experts on the GPU whenever they are activated. However, the benefit of GPU execution depends strongly on the number of tokens processed by each expert.
At large batch sizes, the cost of transferring expert weights can be amortized over more tokens, making GPU execution attractive~\cite{DBLP:conf/asplos/CaoLGSL0GZS25,DBLP:conf/isca/KimWX0YK25}.
In contrast, during low-concurrency inference, each expert typically processes only a small number of tokens, while its weights remain large.
Repeatedly transferring expert weights from CPU memory to the GPU can therefore incur substantial data-movement overhead relative to the computation~\cite{DBLP:conf/ics/YuDZBYWLWL26,DBLP:conf/iclr/KamahoriTGZK25,DBLP:conf/sosp/ChenXZTWDCYLQZO25}.
In such cases, retaining routed experts in CPU memory and executing their FFNs on the CPU can be preferable.
Figure~\ref{fig:intro} illustrates this trade-off: CPU execution remains competitive at small and moderate batch sizes, whereas GPU execution becomes increasingly advantageous as the batch size grows.

This workload-dependent execution behavior defines the scope of our work.
We do not attempt to redesign the CPU--GPU offloading or scheduling policy.
Instead, we focus on the CPU execution path, which remains critical when routed experts are retained in CPU memory under low-concurrency workloads.
Our goal is to make CPU-side expert execution highly efficient and portable, while preserving the standard weight layouts required by existing GPU libraries.
In this way, \SystemName serves as a CPU-side building block for heterogeneous MoE inference: existing or future offloading policies can continue to choose between CPU and GPU execution, while \SystemName provides an efficient CPU backend when CPU execution is selected.

\noindent\textbf{Challenge.} CPU-based MoE inference is constrained by (1) small/irregular matrix multiplications (few tokens per expert) and (2) cross-socket memory traffic when expert weights reside in CPU DRAM.
To exploit matrix engines such as Intel AMX, systems including \kt~\cite{DBLP:conf/sosp/ChenXZTWDCYLQZO25} and Intel Extension for PyTorch (IPEX)~\cite{IPEX} store expert weights in AMX-specific layouts.
While effective for CPU execution, such layouts are incompatible with standard GPU libraries, complicating CPU--GPU offloading and forcing reconversion, duplicated storage, or specialized kernels.
Meanwhile, to mitigate cross-NUMA traffic, \kt further partitions weight tensors across NUMA nodes, requiring deep modifications to the inference framework and its memory allocation behavior.
Together, these choices improve speed but reduce portability and raise the engineering barrier for real deployments.

\noindent\textbf{Our approach.} We present \SystemName, a CPU-optimized library for MoE operators that delivers high performance while remaining \textbf{non-intrusive}: it requires no offline weight reformatting, no framework-level weight sharding, and no custom NUMA-aware allocator (Section~\ref{sec:def-nontru}).
\SystemName preserves general-purpose matrix layouts (row- or column-major) and requires no changes to the framework's weight storage or memory allocation policies; adopting it amounts to replacing a small, well-defined set of operator implementations.

\noindent\textbf{Contributions.} \SystemName achieves these goals via three complementary techniques:
\begin{itemize}
\item \textbf{AMX-efficient micro-kernels for general layouts}.
Based on observations of MoE FFN computation patterns and AMX data layout constraints,
we design micro-kernels that keep expert weights in standard layouts and instead fuse lightweight layout transforms into token gathering and stores (``transform intermediate matrix, not weight matrix”),
enabling efficient AMX utilization without any weight reformatting (Section~\ref{sec:micro}).
\item \textbf{NUMA-aware task partitioning without framework modifications}. Leveraging the fact that expert dimensions in modern MoE models are typically multiples of 128~\cite{MistralAI,qwen,ds3}, we propose a task-partitioning strategy that reduces cross-socket traffic under a simple page-interleaved policy (e.g., \texttt{numactl --interleave=all}) without restructuring weight storage (Section~\ref{sec:cross-NUMA}).
\item \textbf{Attention optimization via MV-to-MM conversion}. During decoding, we convert memory-intensive matrix--vector operations into compute-intensive matrix--matrix execution to better utilize AMX (Section~\ref{sec:attention}).
\end{itemize}

\noindent\textbf{Results.} We evaluate \SystemName on three representative multi-core CPUs: Intel Xeon Gold 6430~\cite{6430}, 6448H~\cite{6448H}, and Max 9462~\cite{9462}.
Across real-world MoE models~\cite{MistralAI,qwen,ds3}, \SystemName consistently outperforms state-of-the-art baselines~\cite{DBLP:conf/sosp/ChenXZTWDCYLQZO25, IPEX, oneDNN}, achieving up to 3.37$\times$ FFN-kernel speedup and up to 2.09$\times$ end-to-end inference speedup.
These results make \SystemName a practical and portable building block for high-performance MoE inference on CPU--GPU nodes.
\section{Background and Related Work}

\subsection{MoE Architecture}

\begin{figure}[!t]
    \centering
    \includegraphics[width=0.52\linewidth]{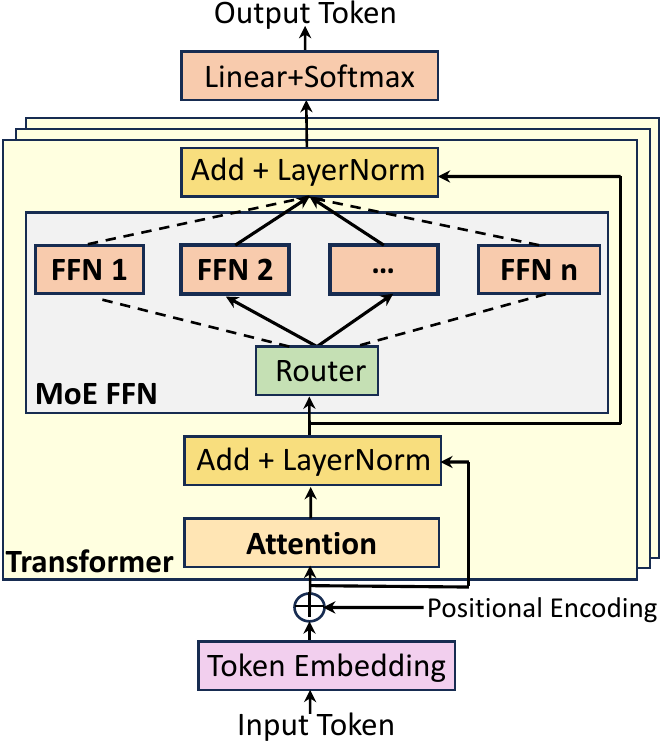}
    \vspace{-1mm}
    \caption{The architecture of MoE model.}
    \label{fig:MoE}
\vspace{-3mm}
\end{figure}

MoE models are built by stacking Transformer layers~\cite{DBLP:conf/asplos/CaoLGSL0GZS25}, where each layer consists of an attention module followed by a feed-forward network (FFN).
In standard dense models (e.g., OPT~\cite{DBLP:journals/corr/abs-2205-01068} and LLaMA-3~\cite{DBLP:journals/corr/abs-2407-21783}), every token is processed by the same FFN. In contrast, an MoE layer replaces this single FFN with multiple independent expert FFNs and uses a gating function to route each token to only a small subset of experts.
As shown in Figure~\ref{fig:MoE}, tokens are first processed by attention and then dispatched to the selected experts, which introduces structured sparsity in both computation and memory traffic.

\subsection{Advanced CPU Compute Unit}\label{sec:AMX}
Modern CPUs integrate matrix engines such as Intel AMX~\cite{AMX} and Arm SME~\cite{SME}, which provide substantially higher throughput than SIMD units (e.g., AVX-512) for matrix workloads.
However, these gains often rely on specific data layouts and incur packing/transform overheads~\cite{DBLP:conf/sosp/ChenXZTWDCYLQZO25}.
In this work, we focus on AMX, which is widely available in current server CPUs (where SME is not yet broadly deployed).

\begin{figure}[ht]
    \centering
    \includegraphics[width=0.86\linewidth]{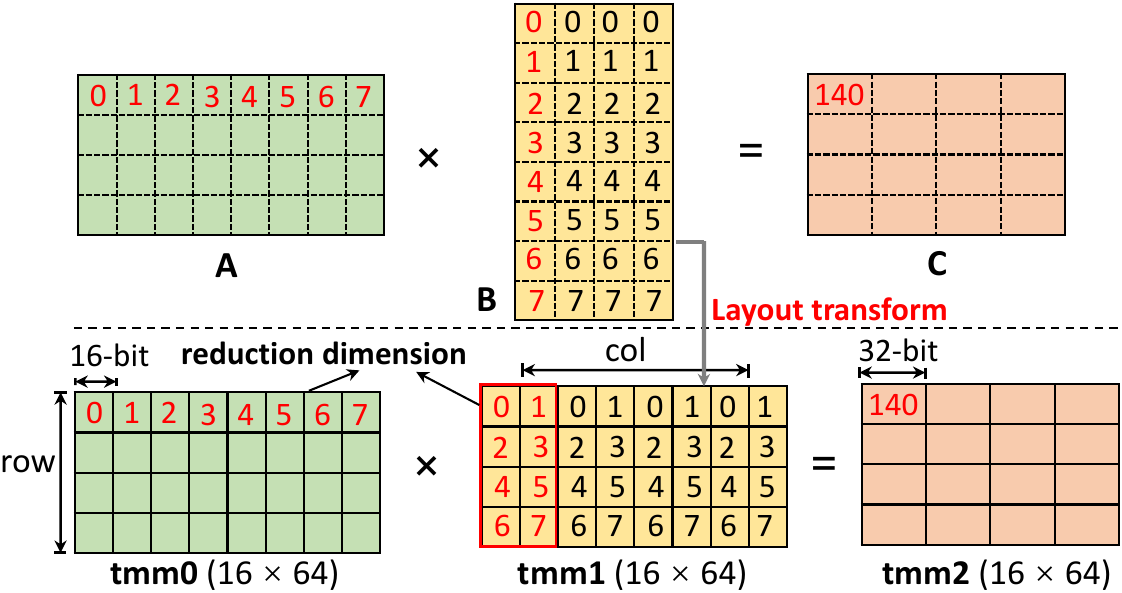}
    \vspace{-1mm}
    \caption{The computational process of AMX’s BF16-precision FMA instructions (all matrices stored in row-major order).}
    \label{fig:AMX}
\vspace{-2mm}
\end{figure}

Figure~\ref{fig:AMX} illustrates the BF16 fused multiply--add (FMA) workflow of AMX. For simplicity, matrices are shown as $4\times 16$ bytes. Each core provides eight tile registers (\tmm{0}$\sim$\tmm{7}), each of size 16$\times$64 bytes.
In the figure, matrix $\mathbf{A}$ matches the layout expected by \tmm{0} (row-major, reduction along columns), which is also a common layout in both CPU and GPU libraries. In contrast, matrix $\mathbf{B}$ does not match the layout required by \tmm{1}$;$ consequently, $\mathbf{B}$ is typically transformed into an AMX-specific layout prior to computation.
We refer to the layout expected by \tmm{0} as the \emph{general layout} and the layout expected by \tmm{1} as the \emph{AMX-specific layout}.

\subsection{NUMA Architecture}

Non-Uniform Memory Access (NUMA) is the dominant architecture for multi-socket servers. In a NUMA system, a core can access memory attached to its local node with higher bandwidth and lower latency than memory attached to a remote node (cross-NUMA access)~\cite{NUMA}. NUMA systems typically expose three memory allocation policies:
\begin{itemize}
    \item \textit{First-touch}: the default OS policy; a page is allocated on the node of the thread that first writes it. This is convenient but can lead to non-deterministic and suboptimal placement.
    \item \textit{Interleaved}: pages are allocated round-robin across all nodes. This policy is deterministic and easy to deploy (e.g., via \texttt{numactl --interleave=all}) without code modifications.
    \item \textit{NUMA-aware}: the application explicitly allocates pages on specific nodes to optimize locality, but this typically requires invasive code changes.
\end{itemize}

\SystemName shows that, when combined with NUMA-aware task partitioning, the simple interleaved policy can mitigate cross-NUMA bottlenecks without modifying the inference framework's weight storage or memory allocator.

\subsection{Related Work} \label{sec:relate}
In resource-constrained scenarios, MoE inference systems typically combine
(i) a high-level policy for placing experts and scheduling transfers across CPU and GPU and
(ii) a low-level implementation of operators (FFN and attention). We therefore organize related work along three axes: \emph{offloading/scheduling}, \emph{CPU operator kernels and data layouts},
and \emph{NUMA locality management}~\cite{DBLP:conf/sosp/ChenXZTWDCYLQZO25,DBLP:conf/dac/ZhongSLWHL25,DBLP:conf/asplos/CaoLGSL0GZS25,DBLP:conf/asplos/FangHHLC0YZ25,DBLP:conf/iclr/KamahoriTGZK25,IPEX,DBLP:conf/isca/KimWX0YK25,DBLP:conf/mlsys/DuWKGX25,DBLP:conf/eurosys/WeiCCMWZY25}.

\subsubsection{Offloading and scheduling}
LIA~\cite{DBLP:conf/isca/KimWX0YK25} adapts offloading decisions based on prompt length, batch size, and inference phase for dense LLMs.
MoE-Lightning~\cite{DBLP:conf/asplos/CaoLGSL0GZS25} pages expert weights to GPU memory and overlaps transfers with GPU FFN execution.
Fiddler~\cite{DBLP:conf/iclr/KamahoriTGZK25} proposes a dynamic offloading strategy and finds that substantial GPU offloading gains occur only beyond a batch-size threshold.
HybriMoE designs prefetching to proactively fetch frequently invoked (``hot'') experts to increase GPU residency~\cite{DBLP:conf/dac/ZhongSLWHL25}. 
\kt~\cite{DBLP:conf/sosp/ChenXZTWDCYLQZO25} uses a static placement strategy, retaining routed experts on the CPU while placing shared experts and attention layers on the GPU.

\cparagraph{Key differentiation.} \SystemName does not propose a new offloading policy. Instead, we focus on making CPU-side operators both fast and portable so they can serve as drop-in building blocks under existing and future policies.

\subsubsection{CPU operator optimization and data layouts}

\kt~\cite{DBLP:conf/sosp/ChenXZTWDCYLQZO25} and IPEX~\cite{IPEX} accelerate CPU-side MoE operators by leveraging AMX,
typically through pre-converting expert weights into AMX-specific layouts (e.g., \tmm{1} in Figure~\ref{fig:AMX}).
T-MAC~\cite{DBLP:conf/eurosys/WeiCCMWZY25} targets low-bit dense LLM inference on embedded CPUs by transforming matrix multiply–accumulate operations into table lookups.

\cparagraph{Why ``layout-first'' can be intrusive.} AMX-specific layouts can improve CPU throughput, but they hinder heterogeneous execution because standard GPU libraries expect general layouts. In practice, this often forces extra weight conversions (AMX-specific layout$\rightarrow$ general layout) or the use of specialized GPU kernels.
Retaining general layouts is therefore essential for seamless CPU–GPU collaborative execution of routed expert FFNs.

\subsubsection{NUMA locality management}
On multi-socket servers, cross-NUMA bandwidth and latency can dominate MoE decoding. \kt reduces remote traffic by partitioning each weight tensor across NUMA nodes and using tensor-parallel execution so each socket primarily accesses local expert weights~\cite{DBLP:conf/sosp/ChenXZTWDCYLQZO25}.


\cparagraph{Key differentiation.} \kt and IPEX are the closest baselines to \SystemName for CPU-side MoE execution with AMX. \SystemName differs in two key ways.
First, instead of requiring AMX-specific \emph{weight} layouts, we preserve general layouts and apply on-the-fly transformations to the intermediate matrix during kernel execution,
maintaining compatibility with standard GPU libraries and CPU--GPU offloading frameworks.
Second, on multi-socket servers,
we avoid framework-level custom NUMA-aware allocators by combining OS-level page interleaving with NUMA-aware task partitioning to reduce cross-socket traffic (Table~\ref{tab:comparison}).

\section{Motivation}


\subsection{The Case for a Non-Intrusive Design} \label{sec:def-nontru}

In local deployments, MoE inference often runs on multi-socket CPU--GPU nodes, where expert parameters reside in CPU DRAM and CPU-side execution becomes performance-critical. Meanwhile, the operator stack must remain easy to integrate across CPU-only, GPU-offload, and hybrid settings~\cite{IPEX, DBLP:conf/isca/KimWX0YK25}. We argue that this requires a \textbf{non-intrusive} design with two principles:
\begin{itemize}
    \item \textbf{General layouts}: expert weights remain in standard row- or column-major formats compatible with CPU and GPU libraries.
    \item \textbf{Framework agnosticism}: no changes to weight storage, tensor semantics, or memory allocation.
\end{itemize}

These principles preserve deployment flexibility: models can move between CPU and GPU without layout conversion or duplication, and NUMA effects can be handled via OS-level mechanisms (e.g., \texttt{numactl --interleave=all}) rather than intrusive changes.

\cparagraph{Non-intrusive criterion.} A method is \emph{non-intrusive} if it requires no (1) offline weight reformatting, (2) framework-level weight sharding/replication changes, or (3) custom NUMA-aware allocators. \SystemName satisfies these by preserving general layouts, using page interleaving, and applying NUMA-aware task partitioning (Table~\ref{tab:comparison}).

Table~\ref{tab:comparison} shows that existing systems trade off performance, portability, and complexity: some retain generality but underutilize CPU performance (e.g., Fiddler), while others achieve high performance via intrusive, hardware-specific optimizations (e.g., IPEX, \kt). \SystemName breaks this trade-off without sacrificing deployability.

\begin{table}[!t]
\centering \footnotesize
\caption{Differentiation from prior work.}
\vspace{-2mm}
\label{tab:comparison}
\begin{tabular}{lccc}
\hline
\textbf{Name} & \textbf{General Layout} & \textbf{\makecell{Cross-NUMA \\Optimization}} & \textbf{Non-Intrusive}\\
\hline
Fiddler~\cite{DBLP:conf/iclr/KamahoriTGZK25} & \cmark & \xmark & \cmark \\
MoE-lightning~\cite{DBLP:conf/asplos/CaoLGSL0GZS25} & \cmark & \xmark & \cmark \\
IPEX~\cite{IPEX} & \xmark & \xmark & \xmark  \\
vLLM~\cite{vllm} & \xmark & \makecell{NUMA-aware \\pre-allocation} & \xmark  \\
\kt~\cite{DBLP:conf/sosp/ChenXZTWDCYLQZO25} & \xmark & \makecell{NUMA-aware \\pre-allocation} & \xmark  \\ 
\textbf{\SystemName} & \cmark & \textbf{interleaved} & \textbf{\cmark}  \\
\hline
\vspace{-4mm}
\end{tabular}
\end{table}


\subsection{Limitations of Previous Works}

To quantify the cost of this trade-off, we evaluate representative systems along two hardware-critical dimensions: (i) AMX utilization and (ii) NUMA locality on multi-socket nodes.


\begin{figure}[ht]
    \centering
    \includegraphics[width=0.6\linewidth]{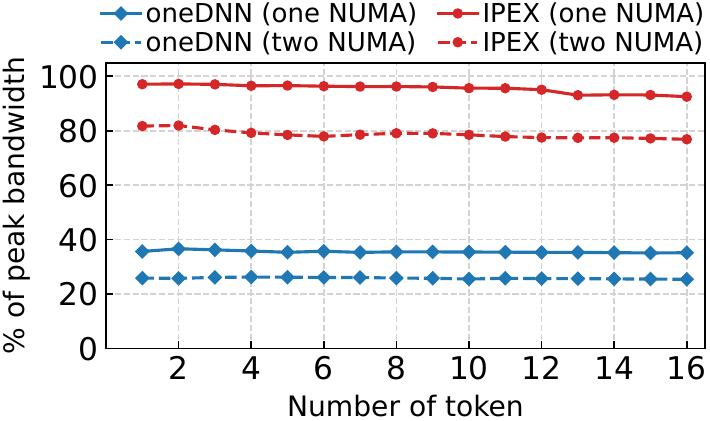}
    \vspace{-1mm}
    \caption{Memory-bandwidth utilization of MoE FFNs in existing methods (IPEX defaults to using TPP). The peak bandwidth refers to the measured peak bandwidth.}
    \label{fig:moti}
\vspace{-2mm}
\end{figure}

\cparagraph{AMX layout overhead.} To satisfy AMX's layout requirements, frameworks typically adopt one of two strategies.
\textit{Runtime conversion} (e.g., oneDNN) stores weights in a general layout and transforms them on the fly during expert execution; this preserves compatibility, but introduces a recurring conversion overhead.
\textit{Pre-conversion} (e.g., IPEX and \kt) stores weights in an AMX-specific layout, maximizing AMX throughput but making the weights incompatible with standard GPU libraries;
CPU--GPU offloading then requires a costly reverse transformation, duplicated weight storage, or bespoke GPU kernels.

\cparagraph{Cross-socket (NUMA) penalty.} Figure~\ref{fig:moti} shows that memory-bandwidth utilization drops by about 20\% when scaling from one to two NUMA nodes. Systems without NUMA-aware execution (e.g., IPEX and oneDNN) incur substantial cross-socket traffic, reducing effective bandwidth available to memory-bound MoE operators. \kt mitigates this by pre-partitioning weights across NUMA nodes to enforce locality, but doing so is intrusive and requires modifying tensor storage as well as the framework's memory allocation behavior.

Overall, current approaches either leave performance untapped or bake in hardware-specific assumptions that reduce portability and increase engineering cost.
This motivates an operator stack that delivers \textbf{high performance without intrusion}.

Motivated by this gap, \SystemName is designed to pursue three goals simultaneously: (1) preserve general matrix layouts to maximize offloading flexibility, (2) fully exploit AMX compute capability,
and (3) avoid cross-NUMA accesses without framework modifications.

\section{Exploiting AMX With General Layouts}\label{sec:micro}

\subsection{Overview and Design Philosophy}
\begin{figure}[!t]
\centering 
\subfloat[][Prior approach\label{fig:exist}]
{ 
\centering
\includegraphics[width=0.95\linewidth]{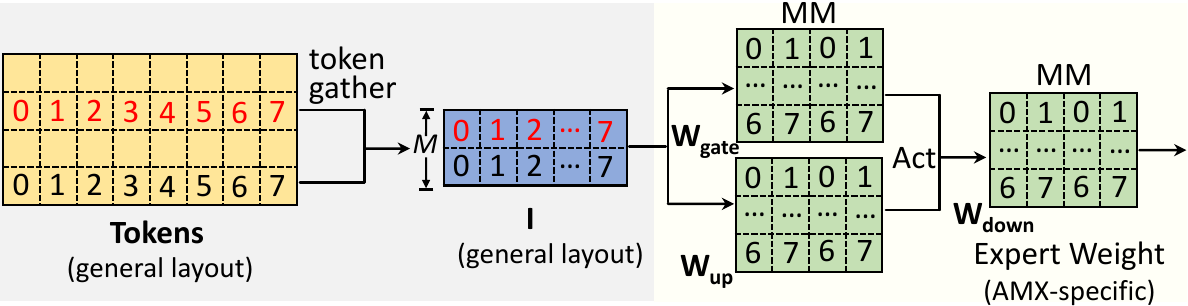}
}\\
\vspace{-1mm}
\subfloat[][Our approach\label{fig:ours}]
{ 
\centering
\includegraphics[width=0.95\linewidth]{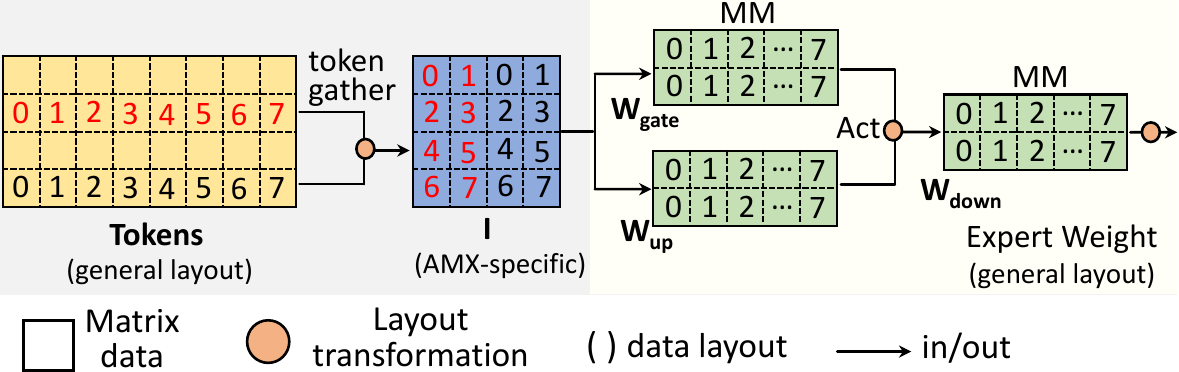}
}
\vspace{-2mm}
\caption{The computation workflow of expert FFN in the MoE.}
\label{fig:expert}
\vspace{-2mm}
\end{figure}

Figure~\ref{fig:expert} illustrates the computation pattern of an expert FFN, where each expert is activated by only a subset of tokens.
To improve efficiency, inference frameworks first \emph{gather} the activated tokens into an intermediate matrix $\mathbf{I}$, so that the expert computation becomes a small number of matrix--matrix multiplications (MM) instead of many matrix--vector multiplications.
Token gathering is a standard step in modern MoE runtimes and is essential for efficient FFN execution on both CPU and GPU~\cite{IPEX,DBLP:conf/sosp/ChenXZTWDCYLQZO25,vllm}.

Given $\mathbf{I}$, the expert FFN computes two projections (with $\mathbf{W_{gate}}$ and $\mathbf{W_{up}}$), applies a nonlinearity (e.g., SiLU), combines the two branches via element-wise multiplication, and then applies $\mathbf{W_{down}}$.
We write the expert FFN as
\begin{equation*}
\small
\mathrm{FFN}(\mathbf{I})= \mathrm{SiLU}(\mathbf{I}\times \mathbf{W_{gate}})\odot(\mathbf{I}\times \mathbf{W_{up}})\times \mathbf{W_{down}} \label{equ:FFN}
\end{equation*}
where $\odot$ denotes element-wise multiplication.

\begin{figure}[!t]
    \centering
    \includegraphics[width=0.86\linewidth]{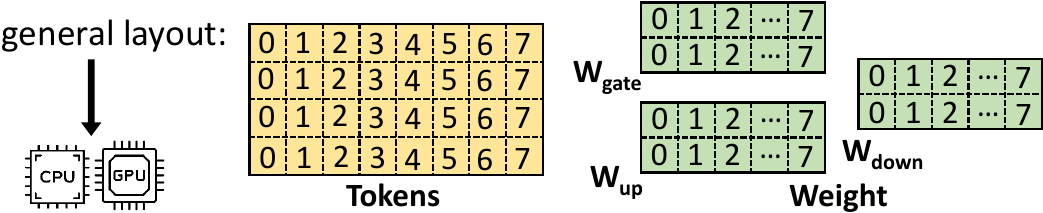}
    \vspace{-2mm}
    \caption{The native data layout of matrices in the inference framework is unified across both CPU and GPU.}
    \label{fig:native}
\vspace{-2mm}
\end{figure}

\noindent\textbf{Why general layouts matter.} In mainstream MoE inference frameworks~\cite{pytorch}, token matrices, expert weights, and output tensors are stored in a unified, general-purpose layout (typically row-major) so the same model state can support both CPU and GPU execution (Figure~\ref{fig:native}).
However, AMX reaches peak throughput only when one MM operand is in an AMX-specific (packed) layout (Section~\ref{sec:AMX}). To avoid per-inference layout conversion, IPEX and \kt pre-convert each expert weight matrix $\mathbf{W}$ into an AMX-specific layout (Figure~\ref{fig:exist}).
While this improves CPU performance, the resulting weight format is incompatible with standard GPU libraries, which makes CPU--GPU offloading inflexible and can force reconversion or special-purpose kernels.

\noindent\textbf{Our approach: transform $\mathbf{I}$, not $\mathbf{W}$.} \SystemName achieves high AMX performance while preserving the framework's general layouts (Figure~\ref{fig:ours}).
This design remains largely unexplored~\cite{IPEX,DBLP:conf/sosp/ChenXZTWDCYLQZO25} due to the unclear interaction between AMX and MoE FFN computation.
Instead of pre-converting weights, we perform lightweight runtime layout transformations on the intermediate matrix $\mathbf{I}$ and fuse them into data movement that the framework already must perform.
Concretely, during token gathering, the selected tokens are loaded into registers and written into $\mathbf{I}$. We fuse a register-level transformation into this path so that $\mathbf{I}$ is produced directly in an AMX-specific layout (the first transformation in Figure~\ref{fig:ours}). The overhead is small because (i) $\mathbf{I}$ is much smaller than the expert weight matrices and (ii) the transform consists of only a handful of register operations.

Specifically, matrix $\mathbf{I}$ corresponds to the tile register labeled \tmm{1} in Figure~\ref{fig:AMX}; this mapping preserves the mathematical form $\mathbf{I} \times \mathbf{W}$.
After computing $\mathbf{I}\times \mathbf{W_{gate}}$ and $\mathbf{I}\times \mathbf{W_{up}}$, we similarly fuse the store path to emit outputs in the layout required by the next AMX MM (the second transformation).
Finally, after multiplying by $\mathbf{W_{down}}$, we apply a register-level transpose before writing the output back in the standard row-major layout (the third transformation).

In summary, we keep expert weights in the general layout (\tmm{0} in Figure~\ref{fig:AMX}) and align $\mathbf{I}$ with the AMX-specific operand (\tmm{1}),
fusing necessary transforms into unavoidable gather/store operations.
This redesign is confined to the FFN operator implementation and requires no changes to the framework's matrix layouts.
\SystemName therefore enables the framework to fully exploit AMX without requiring the framework to be aware of AMX-specific layout constraints.

\vspace{-1mm}
\subsection{Micro-kernel Design} \label{sec:mic}

\subsubsection{Determining Micro-kernel Size}
Expert FFN performance is dominated by its three MM operators, so micro-kernel efficiency is a first-order concern.
We target high efficiency by maximizing the compute-to-memory ratio (CMR) of the micro-kernel~\cite{DBLP:conf/sc/YangFDSW21,DBLP:conf/icpp/WangJZSXY15,DBLP:conf/sc/HeineckeHHP16}, which captures how much arithmetic we can perform per byte loaded from memory.

Many prior designs are primarily tuned for vector registers and fall back to vector units (e.g., AVX-512) when the row dimension of $\mathbf{I}$ is small; for example, \kt uses AMX only when the row dimension exceeds four~\cite{DBLP:conf/sosp/ChenXZTWDCYLQZO25}. In contrast, \SystemName uses AMX for \emph{all} row dimensions and focuses on maximizing reuse within AMX tile registers even in the decoding regime. 

We describe the micro-kernel in BF16, but the same sizing methodology applies to other precisions. Let the micro-kernel tile sizes be $m_r$ (rows), $n_r$ (columns), and $k_r$ (reduction).
AMX tiles support up to 64 bytes along the reduction dimension, corresponding to 32 BF16 elements. Because MoE weight dimensions are typically multiples of 32~\cite{MistralAI,qwen,ds3}, we set $k_r = 32$ to maximize tile utilization (e.g., DeepSeek-V3 uses 7,168 and 2,048; Mixtral-8x7B uses 4,096 and 14,336).

In one micro-kernel step, AMX tiles hold $m_r\cdot 32$ elements of $\mathbf{I}$, $n_r\cdot 32$ elements of $\mathbf{W}$, and $m_r\cdot n_r$ FP32 accumulators. Each core provides eight 1,024-byte tile registers, which constrains $m_r$ and $n_r$ as follows:
\begin{equation}
\small
  \left\{\begin{matrix}
  m_r\cdot 64+n_r\cdot 64+m_r\cdot n_r\cdot 4\leqslant 8\cdot 1024
  \\
  m_r \leqslant M
  \end{matrix}\right.
  \label{equ:cons1}
\end{equation}
where $M$ is the row dimension of $\mathbf{I}$. Each AMX FMA consumes BF16 inputs and accumulates into FP32 (4 bytes), which explains the $\cdot 4$ term.

Each step performs $2\cdot m_r\cdot n_r\cdot 32$ operations (a multiplication and an addition) and loads only $\mathbf{I}$ and $\mathbf{W}$ because the accumulators remain resident in tile registers. Thus,
\begin{equation}
\small
CMR=\frac{2 \cdot m_r \cdot n_r \cdot 32}{m_r\cdot 64+n_r\cdot 64}=\frac{m_r\cdot n_r}{m_r+n_r}
\label{equ:CMR}
\end{equation}
Combining Equations~\ref{equ:cons1} and~\ref{equ:CMR}, the CMR is maximized at $m_r = n_r = 32$. When $M<32$, we set $m_r=M$ to match the available rows.

\begin{algorithm}[htb]\footnotesize
    \caption{Micro-kernel for MM}
    \label{alg:micro}
    \DontPrintSemicolon
    \For{k = 0 $\rightarrow$ $K$ step \ 32} {
        (tmm0 $\sim$ tmm1) \ $\leftarrow$ \textbf{I} ($m$ : $m+32$, $k$ :$k+32$)          \Comment{tileloadd}  \\                        
        (tmm2 $\sim$ tmm3) \ $\leftarrow$ \textbf{W} ($n$ : $n+32$, $k$ :$k+32$)    \Comment{tileloaddt1/tileloadd}         \\
        (tmm4 $\sim$ tmm7) $\leftarrow$ FMA (tmm0$\sim$tmm1, tmm2$\sim$tmm3)                               \\
    }
    Layout transformation  \Comment{the second or third layout transform in Fig.~\ref{fig:ours}}\\
    store (tmm4-tmm7) 
\end{algorithm}

\subsubsection{Implementation Details} 
Algorithm~\ref{alg:micro} outlines the implementation of our AMX MM micro-kernel.
When $M\leq 32$, expert weights exhibit limited temporal reuse in the cache; we therefore load $\mathbf{W}$ using the \load[t1] instruction. \load[t1] performs non-temporal loads that reduce cache pollution and do not trigger hardware prefetching, improving the effective use of main-memory bandwidth under decoding-style access patterns.

The second and third layout transformations in Figure~\ref{fig:ours} are fused into the micro-kernel (line 5 of Algorithm~\ref{alg:micro}).
These transformations execute within registers and the L1 data cache, and therefore introduce negligible overhead, which we also validate experimentally.

\section{Non-Intrusive Cross-NUMA Optimization}\label{sec:cross-NUMA}

\subsection{Performance Bottleneck and Opportunity}
In CPU-centric MoE inference on multi-socket servers, the dominant NUMA bottleneck is remote access to expert weight matrices~\cite{DBLP:conf/sosp/ChenXZTWDCYLQZO25}. In contrast, accesses to the gathered token matrix $\mathbf{I}$ are typically not performance-critical. During prefill, $M$ is determined by batch size, prompt length, and expert activation rate, and in common local deployments the batch size is small (often 1). When prompts are short, $\mathbf{I}$ is much smaller than the expert weights; when prompts are long, prefill becomes compute-bound and is commonly addressed via CPU-GPU or multi-GPU parallelism, which is outside our scope. 
During decoding, $M$ equals the batch size times the activation rate, and $\mathbf{I}$ remains small. 

To avoid cross-socket traffic on weights, prior systems such as \kt use \emph{numa\_alloc\_onnode} to partition each weight matrix across NUMA nodes so that each socket primarily accesses its local portion (Figure~\ref{fig:numa}(a)). However, this approach is intrusive because it requires modifying the framework's tensor storage and memory allocator.

\SystemName instead relies on a standard OS mechanism: interleaving the physical pages of weight matrices across NUMA nodes (e.g., via \texttt{numactl --interleave=all}). With 4~KB pages, each page holds 2,048 BF16 elements; for example, a weight matrix with 6,144 columns occupies three pages per row (Figure~\ref{fig:numa}(b)). Given this page-level interleaving, \SystemName uses task partitioning so that each socket predominantly touches pages local to its NUMA node, while preserving the framework's original weight storage.

\subsection{Task Partitioning Strategy} \label{sec:task}

\begin{figure}[t]
    \centering
    \hspace{-3.2mm}
    \includegraphics[width=1.0\linewidth]{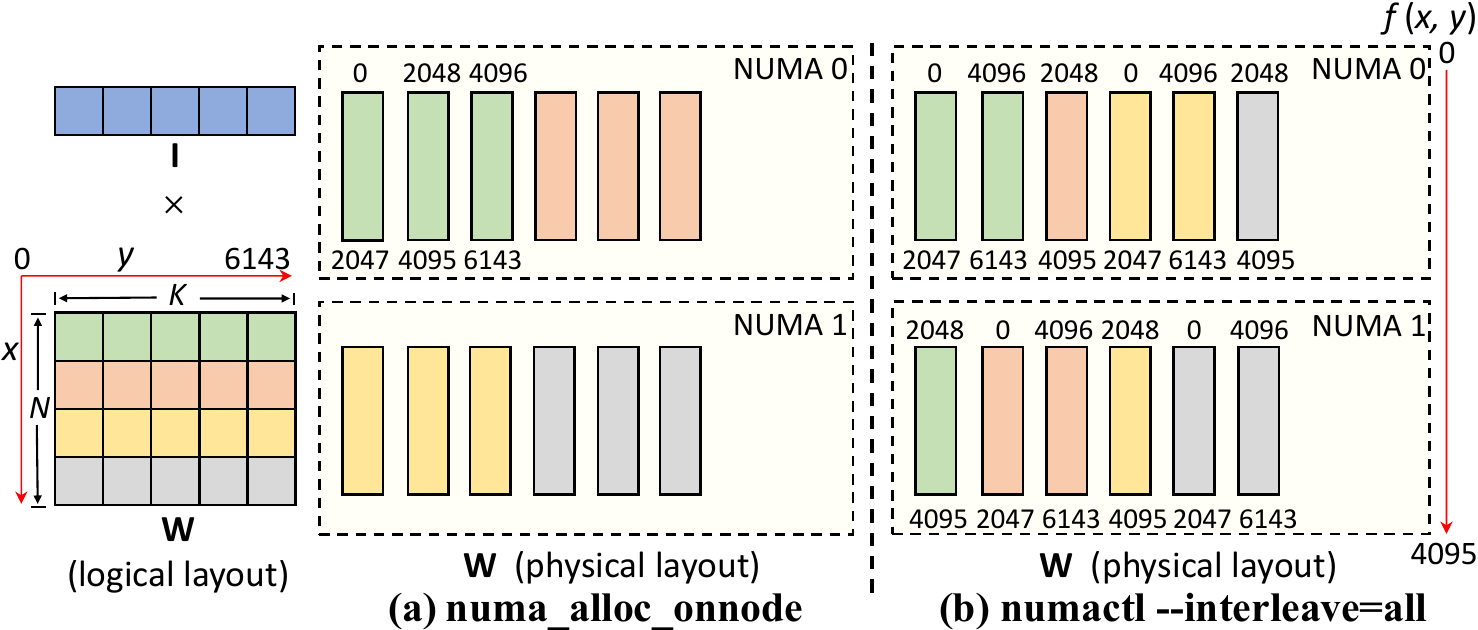}
    \vspace{-2mm}
    \caption{The physical layout of the weight matrix in main memory. The numbers in (a) and (b) denote the coordinates of the elements along the reduction dimension.}
    \label{fig:numa}
\vspace{-2mm}
\end{figure}

Under an interleaved page layout, high MM performance requires two properties: (1) each socket should primarily read weight elements that are local to its NUMA node, and (2) each socket must still be able to assemble contiguous tiles that efficiently feed AMX (in particular, the operand mapped to \tmm{0} in Figure~\ref{fig:AMX}).

\begin{figure}[t]
    \centering
    \includegraphics[width=0.95\linewidth]{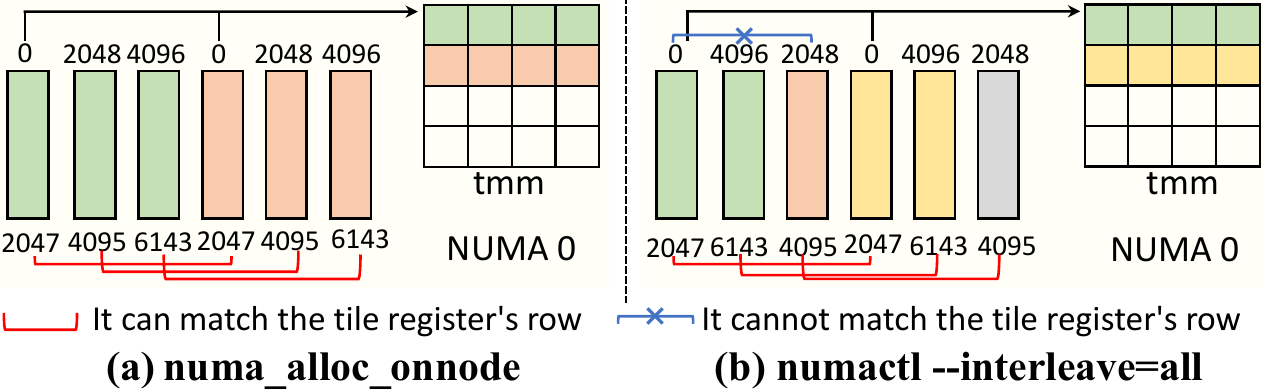}
    \vspace{-1mm}
    \caption{The workflow for mapping data segments from the NUMA 0 (Figure~\ref{fig:numa}) onto tile registers.}
    \label{fig:amx_numa}
\vspace{-2mm}
\end{figure}

With explicit NUMA placement (e.g., \emph{numa\_alloc\_onnode}), these properties are straightforward to satisfy: adjacent matrix rows stored on the same node can be mapped directly to AMX tile rows (Figure~\ref{fig:amx_numa}(a)). Under page interleaving, however, adjacent rows can reside on different NUMA nodes. For example, in Figure~\ref{fig:numa}(b), the first 2,048 elements of row 0 are on NUMA 0, while the first 2,048 elements of row 1 are on NUMA 1.

As a result, as shown in Figure~\ref{fig:amx_numa}(b), on NUMA 0 we may need to combine segments from non-adjacent rows to fill AMX tile rows (e.g., elements $0\sim 2047$ of row 0 and elements $0\sim 2047$ of row 2), while selecting multiple segments within a row to cover the reduction dimension (e.g., elements $0\sim 2047$ and $4096\sim 6143$ of row 0). Next, we show how to systematically identify and organize these segments.

Concretely, we address three subproblems: (1) identifying which weight segments are local to each NUMA node under interleaving; (2) selecting local segments that share the same $y$-range so they can form AMX tile rows; and (3) selecting local segments that share the same $x$ so they provide contiguous ranges along the reduction dimension.
Subproblem (1) enforces locality, while (2) and (3) ensure efficient utilization of AMX.

For ease of description, we define the following terms.

\textbf{$x$ and $y$:} the coordinates of an element in the weight matrix along the row and reduction dimensions, respectively. For example, the labels 0, 2047, and 4096 in Figure~\ref{fig:numa} indicate $y$-coordinates.

\textbf{Data segment $(x,\ y_{start}\sim y_{end})$:} a contiguous range of elements in the same row, from $y_{start}$ to $y_{end}$ along the reduction dimension. For example, the first segment in Figure~\ref{fig:numa}(b) is $(0, 0\sim 2047)$.

\textbf{Function $f(x, y)$:} the page-level ``vertical'' coordinate used to determine the NUMA placement under interleaving. We define
\begin{equation}
\small
f(x,y) = (x\cdot K + y)\bmod (2048\cdot N_{numa}),\ x \in [0,N),\ y \in [0,K)
\label{equ:numa}
\end{equation}
Here, $N$ and $K$ are the row and reduction dimensions of the weight matrix, and $N_{numa}$ is the number of NUMA nodes. The function $f(x,y)$ is periodic with period
\begin{equation}
\small
\begin{split}
T_x &= \frac{2048\cdot N_{numa}}{\gcd(K,2048\cdot N_{numa})}\in [0,N),\\
T_y &= 2048\cdot N_{numa}\in [0,K)
\label{equ:pery}
\end{split}
\end{equation}
where $T_x$, $T_y$ are the periods in $x$ and $y$.
Based on the characteristics of the periodic function, namely $f(x, y) = f(x + T_x, y + T_y)$,
it can be concluded that the data segment $(x, y_{start}\sim y_{end})$ and the data segment $(x+T_x, y_{start}+T_y\sim y_{end}+T_y)$ are guaranteed to reside on the same NUMA node.
Therefore, we only need to identify the data segments within the first period of $x$; the remaining data segments can be derived based on the properties of the periodic function.
All data on each NUMA node must satisfy the following constraints:
\begin{equation}
\small
\begin{split}
id_{numa}\cdot 2048<=f(x,y)<(id_{numa}+1)\cdot 2048\\
\quad id_{numa}\in[0,N_{numa}),x\in[0,T_x) \label{equ:seg}
\end{split}
\end{equation}
$id_{numa}$ denotes the id of a NUMA node.

$T_x$ is usually very small, and this follows from two observations.
First, in modern MoE models, the dimensions of expert weight matrices (i.e., \texttt{hidden\_size} and \texttt{intermediate\_size}) are typically multiples of 128.
This is not accidental: widely used model families such as Qwen~\cite{qwen}, DeepSeek~\cite{ds3}, and Mixtral~\cite{MistralAI} all exhibit this property.
Second, mainstream server platforms most commonly expose two NUMA nodes, while four-node systems are comparatively rare.
For example, in Mixtral-8x7B, $K$ takes values 4,096 or 14,336; on a dual-socket system, Equation~\ref{equ:pery} yields $T_x=1$ and $T_x=2$, respectively.
Similarly, in Qwen3-30B-A3B, $K=2,048$ or 768, corresponding to $T_x=2$ and $T_x=16$.
Because the period $T_x$ is small, we only need to traverse the first period of $x$. For each fixed $x$ in this interval, we can identify the corresponding $y$-ranges.
The lower and upper bounds of each such $y$-range are exactly the $y_{start}$ and $y_{end}$ values of a local data segment.

In summary, the data segments assigned to each NUMA node can be derived in two steps.
First, compute $T_x$ and $T_y$ from the weight shape and the number of NUMA nodes using Equation~\ref{equ:pery}.
Second, determine the data segments for each NUMA node within the first period of $x$ using Equation~\ref{equ:seg}, which solves subproblem (1).
Segments with the same $x$ are then used to assemble contiguous ranges along the reduction dimension of the AMX tile registers, while segments sharing the same $y_{start}$ and $y_{end}$ are used to assemble the tile rows.
This addresses subproblems (2) and (3).
Taking the weight matrix in Figure~\ref{fig:numa}(b) as an example, we have $T_x = 2$ and $T_y = 4096$.
For NUMA 0, when $x=0$ we obtain $y \in [0, 2048)$ and $y \in [4096, 6144)$, and when $x=1$ we obtain $y \in [2048, 4096)$.
Based on these three segments, all segments on NUMA 0 can be derived.
In the implementation, we do not need to explicitly materialize the remaining segments; the kernel can index the weight tensor using only these segments together with $T_x$, which is sufficient to ensure correct and local computation.

Although our design leverages the facts that current MoE models use 128-aligned dimensions and that commercial servers expose only a small number of NUMA nodes, these are not arbitrary assumptions; rather, they arise naturally from both hardware constraints and common model design choices.
Overall, \SystemName enables the framework to minimize cross-NUMA accesses to expert weights without requiring the framework to be aware of NUMA details.

\subsection{Design Details}
To fit the task partitioning in Section~\ref{sec:task}, we make several adjustments to the micro-kernel, as shown in Algorithm~\ref{alg:micro1}.
We use the array $S_{index}$ to store the indices of the data segments assigned to each node, containing only the data segments within a single period of $x$ (i.e., $x<T_x$).
Each entry of the array contains three elements: $x$, $y_{start}$, and $y_{end}$.

Line 1 of Algorithm~\ref{alg:micro1} iterates over one period of $x$.
For a fixed $x$, all local data segments are processed along the reduction dimension, which is implemented by the loop in Line 3.
Specifically, segments that share the same $x$ are multiplied with the corresponding elements of $\mathbf{I}$, and the partial results are accumulated into tile registers \tmm{4}$\sim$\tmm{7} (Lines 2 and 7).
To fully utilize the two tile registers used for the $\mathbf{W}$ operand, 32 segments of $\mathbf{W}$ that share the same $y$ coordinate are loaded from the local NUMA node in each iteration (Line 6), where the $x$ coordinate of adjacent segments differs by $T_x$.
Since the reduction dimension of $\mathbf{W}$ may span multiple NUMA nodes (Figure~\ref{fig:numa}(b)), each node computes a partial result using only its local segments.
After all NUMA nodes finish, the partial results are aggregated to produce the final output.

\begin{algorithm}[htb]\footnotesize 
    \caption{NUMA-optimized micro-kernel for MM}
    \label{alg:micro1}
    \DontPrintSemicolon
    \For{x = 0 $\rightarrow$ $T_x$ step 1 }{
    (tmm4 $\sim$ tmm7) \ $\leftarrow$ 0   \Comment{zeroing for accumulation}\\
    \For{($y_{start}, y_{end}$) = $S_{index}[x] (y_{start}, y_{end})$ $\rightarrow$ $K$ step $T_{y}$ }{
    \For{y = $y_{start}$ $\rightarrow$ $y_{end}$ step \ 32} {
        (tmm0 $\sim$ tmm1) \ $\leftarrow$ \textbf{I} ($m$ : $m+32$, $y$ :$y+32$)         \\
        \tcc{load 32 data segments}
        (tmm2 $\sim$ tmm3) \ $\leftarrow$ \textbf{W} ($x$ : $x+32\cdot T_x$, $y$ : $y+32$)          \\
        (tmm4 $\sim$ tmm7) $\leftarrow$ FMA (tmm0$\sim$tmm1, tmm2$\sim$tmm3)                               \\
    }}}
\end{algorithm}

\vspace{-1mm}
\section{Attention Module Optimization}\label{sec:attention}
The attention module consists of (i) input projections (three matrix multiplications, producing $\mathbf{Q}$, $\mathbf{K}$, and $\mathbf{V}$), (ii) scaled dot-product attention (SDPA), and (iii) an output projection (one matrix multiplication). The projection MMs are optimized by our AMX micro-kernels (Section~\ref{sec:micro}) together with the NUMA-aware task partitioning strategy (Section~\ref{sec:task}), and therefore we focus here on SDPA.
Prior work has largely optimized SDPA for the prefill phase~\cite{DBLP:conf/ics/FuYDS24,dao2022flashattention}, where computation dominates and batching is higher.
We therefore target SDPA optimization in decoding.

\subsection{SDPA Module Optimization}
\begin{figure}[!t]
    \centering
    \includegraphics[width=0.96\linewidth]{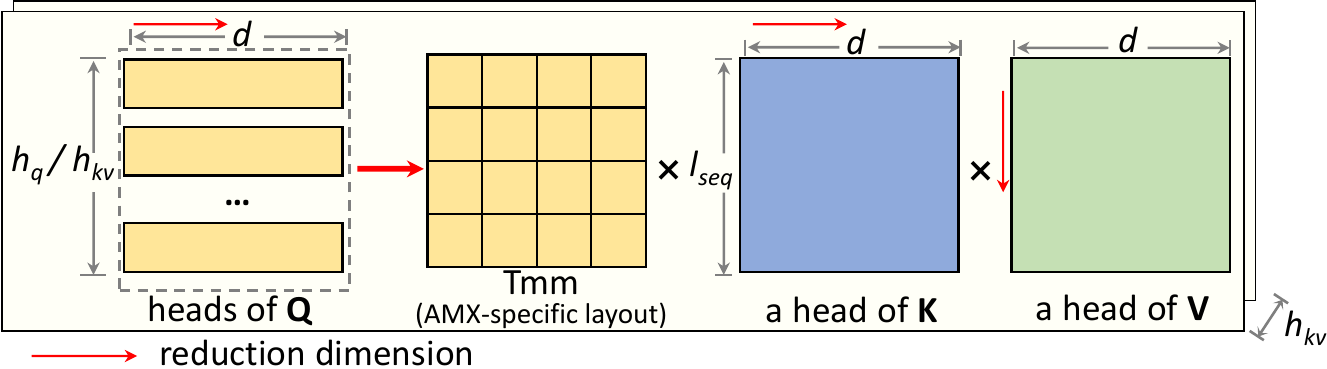}
    \vspace{-2mm}
    \caption{The computational workflow of SDPA during the decoding phase.}
    \label{fig:gqa}
\vspace{-2mm}
\end{figure}

Modern MoE models commonly use grouped-query attention (GQA)~\cite{DBLP:conf/emnlp/AinslieLJZLS23} to reduce KV-cache size while maintaining accuracy (e.g., Qwen and Mixtral). During decoding, the query $\mathbf{Q}$ is split into $h_q$ heads, where each head is a length-$d$ vector. The key $\mathbf{K}$ and value $\mathbf{V}$ are split into $h_{kv}$ heads, where each head is a matrix of size $l_{seq}\times d$ (Figure~\ref{fig:gqa}). Here, $d$ denotes the head dimension and $l_{seq}$ denotes the current sequence length.

In GQA, each key/value head is shared by a group of $\frac{h_q}{h_{kv}}$ query heads. Consequently, computing attention scores for one KV head can be expressed as multiplying a small block of query heads (shape $\frac{h_q}{h_{kv}}\times d$) with the corresponding key head (shape $l_{seq}\times d$), i.e., a MM. The same structure appears again when multiplying attention probabilities by the value head, and thus the decode-phase SDPA kernel naturally exposes a sequence of small MMs.

To exploit AMX in these decode-phase MMs, \SystemName performs lightweight, on-the-fly layout transformations on the small query-head block and keeps the transformed block resident in cache across the subsequent multiplies. Because $\frac{h_q}{h_{kv}}$ is small (e.g., 8 in Qwen3-30B-A3B~\cite{qwen}), the transformation cost is negligible relative to the MM work. For the subsequent multiplication with $\mathbf{V}$, \SystemName similarly converts the accessed value-head tiles into an AMX-specific layout on demand, enabling high AMX utilization without changing the framework's native tensor layouts.

\vspace{-1mm}
\section{Experimental Setup}

\subsection{Hardware Details}
We evaluate \SystemName on two modern multi-socket server platforms: a dual-socket Intel Xeon Gold 6430 (\textbf{6430})~\cite{6430} and a quad-socket Intel Xeon Gold 6448H (\textbf{6448H})~\cite{6448H} (Table~\ref{tab:hardware}). Memory bandwidth is measured using STREAM~\cite{NUMA}. The local (per-socket) bandwidth and cross-socket bandwidth quantify the cost of NUMA-remote accesses.
Although both CPUs support eight memory channels per socket, our 6448H machine is populated with only six memory modules per socket, resulting in lower measured local bandwidth than the fully populated 6430. The 6430 system also includes an NVIDIA RTX A6000 GPU~\cite{A6000} for hybrid CPU--GPU experiments~\cite{DBLP:conf/sosp/ChenXZTWDCYLQZO25}, whereas the 6448H system is CPU-only.


\begin{table}[t]
\centering
\small
    \caption{\textbf{Main characteristics of each platform}}
    \vspace{-3mm}
    \begin{tabular}[H]{lcc}
        \toprule
        & \textbf{6430} & \textbf{6448H}  \\
        \midrule
        \textbf{Peak BF16 TFLOPS}   & 134.4  & 300.4 \\
        \rowcolor{gray!10} \textbf{Cores}  & 32x2  & 32x4 \\
        \textbf{Frequency}  & 2.1 GHz  & 2.4 GHz  \\
        \rowcolor{gray!10} Number of NUMA nodes  & 2  & 4  \\
                           Local Bandwidth   & 234 GB/s   & 158 GB/s \\
        \rowcolor{gray!10} Cross-NUMA Bandwidth & 95 GB/s & 32 GB/s  \\
        \bottomrule
    \end{tabular}
\label{tab:hardware}
\vspace{-1mm}
\end{table}

\subsection{Competitive Baselines}
We compare \SystemName against four widely used CPU(-centric) inference stacks.
All baseline configurations follow the official recommendations.

\noindent\textbf{IPEX~\cite{IPEX}:} Intel Extension for PyTorch, which optimizes PyTorch execution on Intel CPUs. By default, it uses Tensor Processing Primitives (TPP)~\cite{DBLP:conf/sc/GeorganasKAAABB21,DBLP:journals/corr/abs-2511-13764}.

\noindent\textbf{\kt~\cite{DBLP:conf/sosp/ChenXZTWDCYLQZO25}:} A framework for locally deployed MoE models with static CPU--GPU offloading: routed experts run on the CPU, while attention and shared experts run on the GPU. We use the \texttt{SOSP'25} release.

\noindent\textbf{vLLM~\cite{vllm}:} A widely used LLM serving framework; we use its CPU-only variant~\cite{vllm_cpu}.

\noindent\textbf{oneDNN~\cite{oneDNN}:} Intel's high-performance DNN primitive library; we use oneDNN \texttt{3.10.1}.

\subsection{Workloads}

We evaluate \SystemName at two levels: \textbf{single-operator (FFN) kernels} and \textbf{end-to-end} model inference.

\cparagraph{FFN Kernel Evaluation.} We extract the expert-matrix shapes from four MoE models: Mixtral-8x7B, Mixtral-8x22B~\cite{MistralAI}, DeepSeek-V3~\cite{ds3}, and Qwen3-235B-A22B~\cite{qwen}. Their expert dimensions and routing configurations are summarized in Table~\ref{tab:models}. We report the geometric mean over 100 runs.

\cparagraph{End-to-End Inference Evaluation.} We measure end-to-end inference latency for Mixtral-8x7B, Qwen3-30B-A3B, and Qwen2-57B-A14B~\cite{qwen2}. We integrate \SystemName by replacing the corresponding CPU operators in IPEX and \kt, yielding two configurations: \textit{\SystemName (IPEX)} and \textit{\SystemName (KTrans)}. Unless otherwise stated, we use batch size 1 to reflect low-concurrency, latency-oriented deployments and to match prior work~\cite{DBLP:conf/sosp/ChenXZTWDCYLQZO25,DBLP:conf/eurosys/WeiCCMWZY25}. We report the geometric mean over 5 runs.

\begin{figure*}[!t]
\centering 
\subfloat[][On 6430\label{fig:6430}]
{ 
\centering
\includegraphics[width=0.92\linewidth]{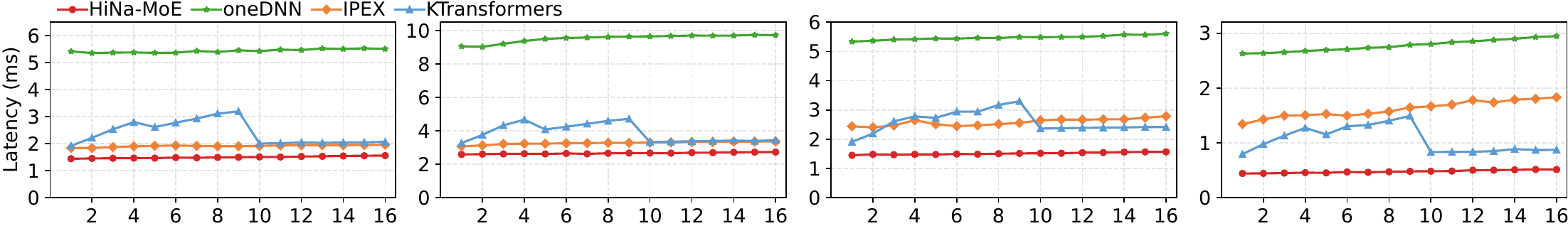}
}\\
\vspace{-1mm}
\subfloat[][On 6448H\label{fig:6448H}]
{ 
\centering
\includegraphics[width=0.92\linewidth]{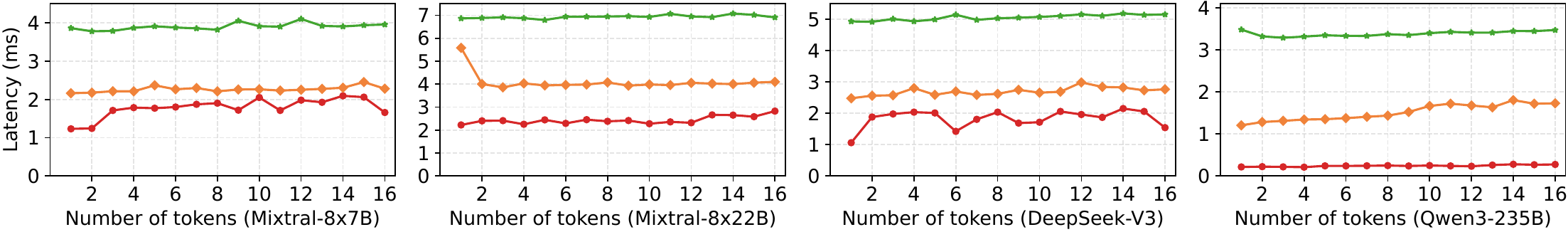}
}
\vspace{-1mm}
\caption{Performance of the FFN operator on the 6430 and 6448H platforms. The X-axis represents the number of tokens assigned to each expert, and the Y-axis represents the latency (in ms) of the MoE FFN.}
\label{fig:FFN_cpu}
\vspace{-2mm}
\end{figure*}

\section{Evaluation Results}\label{sec:eval_results}

\subsection{FFN Kernel Performance}
We begin by evaluating the core MoE FFN operator under varying per-expert token counts, using the exact expert matrix shapes of real-world MoE models (Table~\ref{tab:models}).

\begin{table}[t]
\fontsize{7.5}{9}\selectfont
    \caption{Expert weight matrix dimensions and routing strategies of evaluated MoE models}
    \vspace{-3mm}
    \begin{tabular}[H]{lccc}
        \toprule
        &  \textbf{Gate, Up} & \textbf{Down} & \textbf{Routing Strategy}\\
        \midrule
           Mixtral-8x7B&$4,096\times 14,336$ &$14,336\times 4,096$ & Top-2\\
           Mixtral-8x22B&$6,144\times 16,384$ &$16,384\times 6,144$ & Top-2\\
           DeepSeek-V3&$7,168\times 2,048$ &$2,048\times 7,168$ & Top-8\\
           Qwen3-235B &$4,096\times 1,536$ &$1,536\times 4,096$ & Top-8\\
        \bottomrule
    \end{tabular}
\label{tab:models}
\vspace{-1mm}
\end{table}

Figure~\ref{fig:FFN_cpu} shows that \SystemName consistently outperforms all baselines. 
Averaged across all workloads, it achieves 1.73$\times$ and 1.68$\times$ speedups over IPEX and \kt, respectively, with maximum observed speedups of 3.37$\times$ and 2.16$\times$.
On the 6430, \SystemName reaches up to 96\% of the measured peak local memory bandwidth, suggesting that performance is close to the platform's bandwidth ceiling. These gains come from (i) NUMA-aware task partitioning that avoids remote weight traffic and (ii) AMX micro-kernels that sustain high utilization while preserving general-purpose layouts.
In contrast, oneDNN is slower due to the need to convert the weight matrices into an AMX-specific layout.

MoE FFN latency is largely insensitive to small changes in per-expert token count (e.g., 1$\sim$16 tokens).
The dominant cost is streaming large expert weights from memory, whereas the gathered-token matrix $\mathbf{I}$ contributes little to the total traffic.
\kt shows a slight throughput drop around 4 tokens, as it uses AMX only when an expert processes more than 4 tokens; otherwise, it falls back to AVX-512~\cite{DBLP:conf/sosp/ChenXZTWDCYLQZO25}.
\SystemName uses AMX efficiently even for single-token decoding.

\SystemName's advantage is more pronounced on the 6448H (4 sockets) than on the 6430 (2 sockets) because the 6448H has much lower cross-socket bandwidth (32~GB/s vs. 95~GB/s). Avoiding remote weight traffic therefore yields larger relative benefits as NUMA topologies become more complex.

\begin{figure}[!t]
    \centering
    \includegraphics[width=0.8\linewidth]{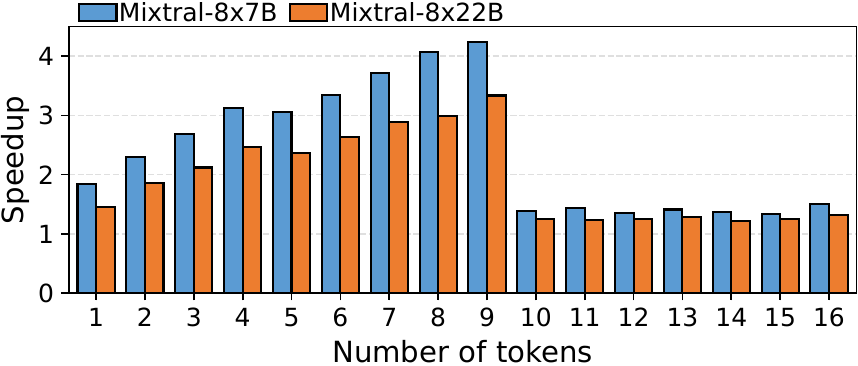}
    \vspace{-3mm}
    \caption{Speedup of \SystemName over \kt for the INT8-quantized FFN kernel on the dual-socket 6430 platform.}
    \label{fig:int8}
\vspace{-1mm}
\end{figure}

We further evaluate the INT8-quantized FFN kernel. For a fair comparison, both \SystemName and \kt use the same symmetric group-wise quantization scheme.
As shown in Figure~\ref{fig:int8}, \SystemName achieves an average speedup of 2.18$\times$ over \kt, demonstrating the effectiveness of our approach under INT8 precision as well.

\subsection{End-to-end Inference Performance}
\begin{figure}[!t]
    \centering
    \includegraphics[width=0.72\linewidth]{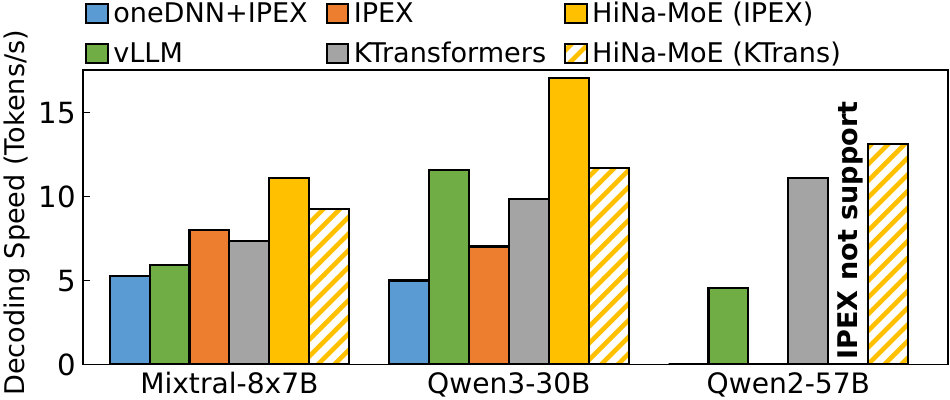}
    \vspace{-3mm}
    \caption{Comparison of decoding speed between \SystemName and the state-of-the-art baselines.}
    \label{fig:decode}
\vspace{-1mm}
\end{figure}

\begin{figure}[!t]
    \centering
    \includegraphics[width=0.72\linewidth]{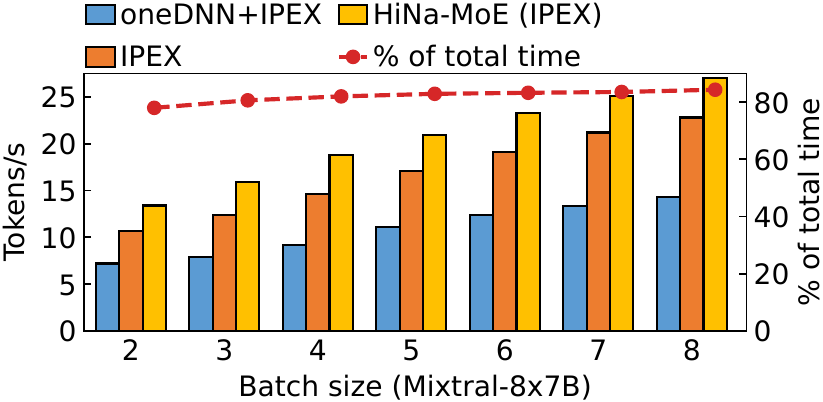}
    \vspace{-3mm}
    \caption{Comparison of batch decoding performance (left: throughput; right: \% of total runtime for optimized FFNs).}
    \label{fig:batch_decode}
\vspace{-1mm}
\end{figure}

\cparagraph{Decoding Phase.}We first evaluate end-to-end decoding performance for Mixtral-8x7B, Qwen3-30B-A3B, and Qwen2-57B-A14B with batch size 1, using 128 input and output tokens on a dual-socket 6430 CPU + RTX A6000 system.
Figure~\ref{fig:decode} shows that \SystemName achieves the highest decoding throughput, with average speedups of 1.99$\times$, 1.70$\times$, and 1.22$\times$ over vLLM (CPU-only), IPEX, and \kt, respectively, and maximum speedups of 2.87$\times$, 2.09$\times$, and 1.25$\times$.
For IPEX and \kt, we use their \SystemName-integrated counterparts to isolate the impact of the CPU operator implementation.
These gains are consistent with our FFN-kernel analysis: decoding is dominated by streaming expert weights from memory, making memory bandwidth and kernel efficiency critical.
\SystemName reduces cross-socket traffic through NUMA-aware task partitioning and improves kernel efficiency for small per-expert token counts compared with IPEX and \kt, respectively.

The results also highlight the benefit of preserving a general-purpose weight layout.
Both oneDNN and \SystemName keep expert weights in the general layout, but oneDNN performs weight-layout transformation at runtime for each inference, whereas \SystemName fuses lightweight transformations into the existing token gather/store paths.
Thus, \SystemName avoids the recurring transformation overhead while retaining compatibility with standard CPU--GPU execution.

To examine whether these benefits persist beyond batch size 1, Figure~\ref{fig:batch_decode} further reports decoding performance across different batch sizes.
\SystemName consistently outperforms IPEX over the evaluated range.
This is expected because expert FFNs account for approximately 80\% of the decoding runtime, making the optimized FFN execution the dominant contributor to end-to-end performance.

\cparagraph{Prefill Phase.}We next evaluate long-context prefill by integrating different CPU operator backends into SGLang (SGL)~\cite{DBLP:conf/nips/ZhengYXS0YCKSGB24} on the 6430 CPU + A6000 GPU system, using Qwen2-57B-A14B with batch size 1. During prefill, GQA attention runs on the GPU, while expert FFNs primarily run on the CPU. Table~\ref{tab:prefill} reports the prefill throughput across input sequence lengths.

Without GPU scheduling, SGL+\SystemName achieves performance comparable to SGL+\kt, showing that preserving the general-purpose weight layout does not sacrifice CPU-side execution efficiency.
More importantly, this general layout also allows GPU scheduling without requiring expert-weight reconversion.
We emphasize that we do not introduce a new scheduling policy; rather, \SystemName's layout compatibility enables SGLang to schedule expert execution on the GPU while retaining the same CPU backend.
With GPU scheduling, SGL+\SystemName consistently outperforms SGL+\kt across all evaluated sequence lengths, achieving speedups from 1.29$\times$ at 2K to 1.67$\times$ at 32K.
These results demonstrate that \SystemName provides both efficient CPU execution and the layout flexibility required for heterogeneous CPU--GPU inference.

\begin{table}[!t]
\centering
\fontsize{7.5}{9}\selectfont
\caption{Prefill Throughput on Qwen2-57B-A14B (tokens/s)}
\vspace{-3mm}
\label{tab:prefill}
\begin{tabular}{c|c|c|c}
\hline
\makecell{\textbf{Input}\\\textbf{Length}} & 
\makecell{SGL + \SystemName\\+ GPU Scheduling} & 
\makecell{SGL + \SystemName} & 
\makecell{SGL\\+ \kt} \\
\hline
2K  & 1526.66 & 1293.36 & 1187.43 \\
4K  & 2065.21 & 1348.27 & 1317.39 \\
8K  & 1977.82 & 1357.80 & 1377.77 \\
16K & 2134.25 & 1221.91 & 1343.52 \\
32K & 2085.47 & 1187.70 & 1252.54 \\
\hline
\end{tabular}
\vspace{-1mm}
\end{table}

Both SGL+\SystemName and SGL+\kt exhibit a similar trend as the sequence length increases: their prefill throughput improves from 2K to 8K, peaks at 8K, and gradually decreases at 16K and 32K. This trend arises from the different scaling behaviors of GPU-side GQA attention and CPU-side expert FFN. As shown in Table~\ref{tab:operator-performance}, increasing the sequence length from 2K to 8K improves the effective computational performance of both operators, with attention increasing from 97.88 to 119.97~TFLOPS and FFN from 13.44 to 15.86~TFLOPS, indicating improved hardware utilization.
Beyond 8K, however, the performance gains become marginal as both operators approach saturation. Meanwhile, GQA attention grows quadratically with the sequence length ($O(l^2)$), whereas the expert FFN grows approximately linearly.
As a result, the attention latency increases rapidly from 8.02~ms at 8K to 85.466~ms at 32K despite nearly saturated execution performance. Consequently, the rapidly increasing attention cost can no longer be offset by improved execution efficiency.
Although FFN remains the dominant component, the growing attention overhead reduces the benefit of processing longer sequences, causing prefill throughput to decrease.

\begin{table}[t]
\centering
\fontsize{7.5}{9}\selectfont
\caption{Effective computational performance of GPU-side attention and CPU-side expert FFN in SGL+\kt across different input sequence lengths.}
\vspace{-3mm}
\label{tab:operator-performance}
\begin{tabular}{c|cc|cc}
\hline
\multirow{2}{*}{\textbf{Input Length}} &
\multicolumn{2}{c|}{\textbf{Attention (GPU)}} &
\multicolumn{2}{c}{\textbf{Expert FFN (CPU)}} \\
\cline{2-3}\cline{4-5}
& Time (ms) & TFLOPS & Time (ms) & TFLOPS \\
\hline
2K  & 1.536  & 97.88  & 67.124  & 13.44 \\
4K  & 3.224  & 111.91 & 119.981 & 15.04 \\
8K  & 8.020  & 119.97 & 227.537 & 15.86 \\
16K & 23.021 & 125.38 & 452.798 & 15.93 \\
32K & 85.466 & 112.57 & 919.989 & 15.69 \\
\hline
\end{tabular}
\vspace{-1mm}
\end{table}

\subsection{Ablation Study}
\begin{figure}[!t]
    \centering
    \includegraphics[width=0.6\linewidth]{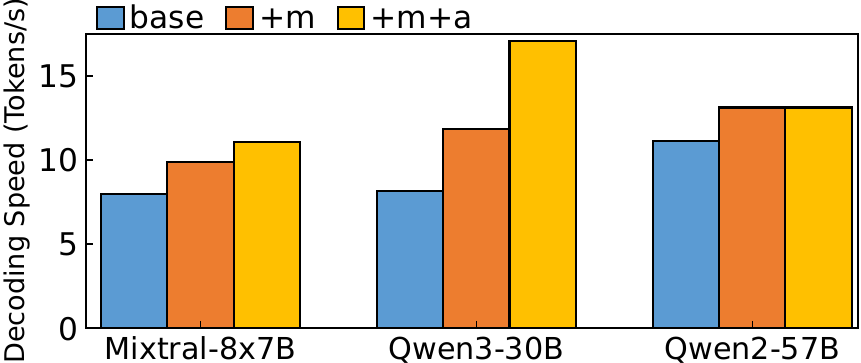}
    \vspace{-2mm}
    \caption{Ablation study on the decoding phase.}
    \label{fig:Ablation}
\vspace{-2mm}
\end{figure}

We analyze the contribution of each component: the micro-kernel (+m), including FFN optimizations for both the general layout and page-interleaving, and the decode-phase attention optimization (+a).
Figure~\ref{fig:Ablation} reports the decoding throughput improvement brought by each component at a batch size of 1, with IPEX serving as the baseline for Mixtral-8x7B and Qwen3-30B-A3B, and \kt for Qwen2-57B-A14B.

Both optimizations improve performance, with micro-kernel optimization (+m) providing the largest gain and improving performance by 27\% on average over the baseline.
For Qwen2-57B-A14B, attention optimization provides no additional benefit because \kt offloads attention computation to the GPU, whereas (+a) targets the CPU-side attention path.
In contrast, Qwen3-30B-A3B exhibits a more pronounced benefit from (+a), as its attention accounts for a larger fraction of the decoding computation than in Mixtral-8x7B, making CPU-side attention optimization more impactful on end-to-end performance. This result highlights that the effectiveness of attention optimization depends on both the relative cost of attention and the underlying offloading strategy.

\subsection{Cross-NUMA Data Traffic Analysis}
\begin{figure}[!t]
    \centering
    \includegraphics[width=0.9\linewidth]{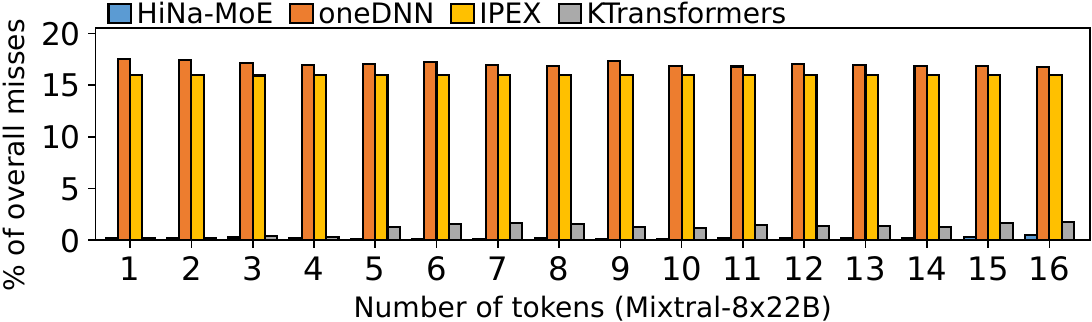}
    \vspace{-2mm}
    \caption{The proportion of remote L3 cache misses to overall L3 cache misses.}
    \label{fig:traffic}
\vspace{-2mm}
\end{figure}

Using Linux \texttt{perf}, we quantify cross-socket traffic by measuring the fraction of L3-miss loads served by remote DRAM (\texttt{mem\_load\_l3\_miss\_retired.remote\_dram}) among all DRAM-served L3-miss loads. Figure~\ref{fig:traffic} shows that \SystemName's remote-access fraction is nearly zero, validating that our task partitioning (Section~\ref{sec:cross-NUMA}) largely eliminates remote accesses to weights. \kt also shows low remote traffic due to its manual weight partitioning.
While oneDNN and IPEX exhibit similar remote fractions, oneDNN additionally pays runtime overhead for weight-layout transformations, leading to worse overall performance.

\begin{figure}[!t]
    \centering
    \includegraphics[width=0.9\linewidth]{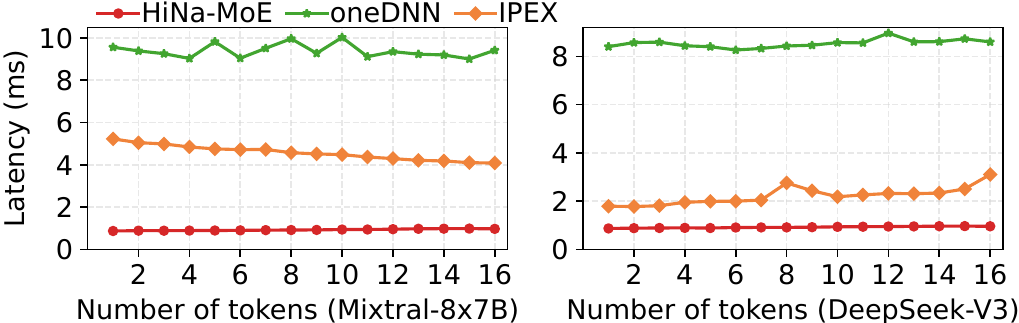}
    \vspace{-2mm}
    \caption{Performance of the MoE FFN on the 9462 CPU.}
    \label{fig:hbm}
\vspace{-2mm}
\end{figure}
\subsection{Evaluation on an HBM-based CPU}\label{sec:hbm}
We evaluate FFN-kernel performance on a dual-socket Intel Xeon CPU Max 9462 (\textbf{9462})~\cite{9462}. Each socket provides 64~GB of HBM (HBM-only mode). The measured peak bandwidth is 853~GB/s; local NUMA accesses reach 425~GB/s, while cross-socket accesses drop to 89~GB/s.

Because HBM capacity is insufficient for full end-to-end MoE inference on our evaluated models, we report FFN-kernel results only.
Figure~\ref{fig:hbm} shows that \SystemName outperforms oneDNN and IPEX by 9.65$\times$ and 3.41$\times$ on average. The larger gap on 9462 is explained by the higher penalty of remote accesses: the local-to-remote bandwidth ratio is $\sim$5$\times$ on 9462 versus $\sim$2.3$\times$ on 6430. As the cost of cross-socket accesses increases, \SystemName's NUMA-aware scheduling yields correspondingly larger benefits.

\subsection{Integrating \SystemName into MoE Inference Frameworks}

\begin{figure}[ht]
    \centering
    \includegraphics[width=0.85\linewidth]{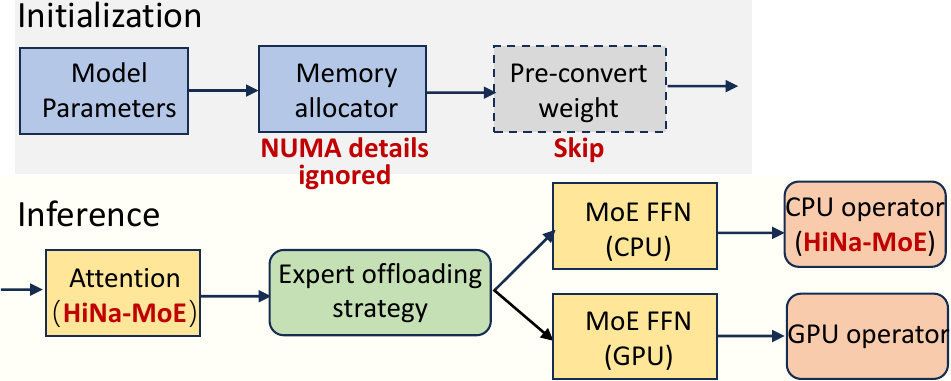}
    \vspace{-2mm}
    \caption{Workflow of the MoE inference framework.}
    \label{fig:inte}
\vspace{-2mm}
\end{figure}

Figure~\ref{fig:inte} summarizes the workflow of existing MoE inference frameworks. Integrating \SystemName requires only two changes:
(1) skip the weight pre-conversion step used by implementations based on the AMX-specific layout, and (2) replace the framework's CPU-side attention and FFN operators with \SystemName implementations.
For example, in IPEX we replace the \emph{mixtral\_moe\_tpp\_kernl\_impl} function responsible for FFN computation; in total, integration requires modifying about 30 lines of framework code.

\section{Discussion and Future Work}
Our evaluation shows that \SystemName achieves consistent speedups across multi-socket servers and real-world MoE models. While the current implementation targets latency-oriented inference at small batch sizes on AMX-capable CPUs, the underlying design principles generalize.

\cparagraph{Adaptation to other CPU architectures.} \SystemName's core idea---preserve general-purpose weight layouts and adapt the computation kernels---extends beyond AMX. Prior work suggests that Arm's SME can support tile-oriented micro-kernels similar in spirit to AMX~\cite{DBLP:conf/ipps/DengYFD26}. A natural direction is to apply the same ``transform intermediate matrix $\mathbf{I}$, not $\mathbf{W}$'' philosophy to SME by fusing lightweight layout transformations into token gathering and kernel execution, with tile sizes chosen to match the available register space and instruction semantics.

\cparagraph{Multi-batch and CPU--GPU collaborative inference.} For larger batch sizes and long-prompt prefill, CPU-only routed-expert execution can become a bottleneck because GPU compute and bandwidth are typically higher~\cite{DBLP:conf/asplos/CaoLGSL0GZS25,DBLP:conf/isca/KimWX0YK25}. A promising direction is dynamic expert offloading based on prompt length, batch size, and available GPU memory, overlapping transfers with CPU computation~\cite{DBLP:conf/ics/YuDZBYWLWL26}. Because \SystemName preserves general-purpose layouts and avoids framework-level allocator changes, experts can move between CPU and GPU without layout-conversion overhead, making it a natural substrate for such schedulers.

\section{Conclusion}
We have presented \SystemName, a high-performance, non-intrusive operator library for MoE inference on multi-socket, multi-core CPUs. It preserves general-purpose matrix layouts for compatibility with CPU--GPU offloading frameworks, while improving performance through (1) AMX-efficient micro-kernels that fuse token gathering with lightweight, on-the-fly data movement, (2) NUMA-aware task partitioning that exploits the periodicity of interleaved pages without modifying the framework allocator, and (3) decode-phase attention optimization that converts memory-bound matrix--vector operations into compute-intensive matrix--matrix multiplications.

Across multiple server-class CPUs and real-world MoE models, \SystemName achieves up to 3.37$\times$ FFN-kernel speedup and up to 2.09$\times$ end-to-end MoE inference speedup over state-of-the-art
baselines, while maintaining near-zero cross-socket traffic and plug-and-play integration.
\begin{acks}

We thank Guangnan Feng and Zheng Wang for their helpful discussions.
This work is supported by the National Natural Science Foundation of China under Grant No.62421002 and
No.U24B20151, and the Hunan Provincial Natural Science Foundation of China under Grant No. 2026JJ20082.
\end{acks}

\bibliographystyle{ACM-Reference-Format}
\balance
\bibliography{ref}

\end{sloppypar}
\end{document}